\documentclass[useAMS,usegraphicx]{mn2e}
\usepackage{epsfig}
\newcommand{\Nmm} {$N_\mathrm{mm}$}
\newcommand{\Nj} {$N_\mathrm{Jeans}$}

\newcommand{\ndiam} 	{$n_\mathrm{0.1\,pc}$}

\newcommand{\kms}   {km~s$^{-1}$}

\newcommand{\cmt}   {cm$^{-3}$}
    
\newcommand{\lo}    {$L_{\sun}$}
\newcommand{\mo}    {$M_{\sun}$}

\newcommand{\nh}    {NH$_3$}
\newcommand{\nth}   {N$_2$H$^+$}

\newcommand{\eg}    {e.\,g.,}
\newcommand{\ie}    {i.\,e.,}

\newcommand{\supa}  {$^\mathrm{a}$}
\newcommand{\supb}  {$^\mathrm{b}$}
\newcommand{\supc}  {$^\mathrm{c}$}
\newcommand{\supd}  {$^\mathrm{d}$}

\def\ltsima{$\; \buildrel < \over \sim \;$}    
\def\lesssim{\lower.5ex\hbox{\ltsima}}           
\def\gtsima{$\; \buildrel > \over \sim \;$}    
\def\gtrsim{\lower.5ex\hbox{\gtsima}}           

\def\apj{ApJ}
\def\apjl{ApJL}
\def\aap{A\& A}

\def\mnras{MNRAS}

\def\mnras{{MNRAS}}
\def\pasj{{PASJ}}

\title[Evidence of pure thermal Jeans fragmentation]{Gravity or turbulence? --III. Evidence of pure thermal Jeans fragmentation at $\sim0.1$~pc scale
} \author[Palau et al.]{Aina  Palau$^1$\thanks{E-mail:a.palau@crya.unam.mx}, Javier Ballesteros-Paredes$^1$, Enrique V\'azquez-Semadeni$^{1}$, 
  \newauthor \'Alvaro S\'anchez-Monge$^2$, Robert Estalella$^{3}$\thanks{The ICC (UB) is a CSIC-Associated Unit through the ICE (CSIC).}, S. Michael Fall$^{4}$, Luis A. Zapata$^{1}$, 
  \newauthor  Vianey Camacho$^1$, Laura G\'omez$^{5, 6}$, Ra\'ul Naranjo-Romero$^1$, Gemma Busquet$^{7}$,
  \newauthor Francesco Fontani$^8$ \\
  $^{1}$ Instituto de Radioastronom\'ia y Astrof\'isica, Universidad
  Nacional Aut\'onoma de M\'exico, P.O. Box 3-72, 58090 Morelia, Michoac\'an, M\'exico\\ 
  $^{2}$ I. Physikalisches Institut der Universit\"at zu K\"oln, Z\"ulpicher Strasse 77, 50937 K\"oln, Germany\\
  $^{3}$ Departament d'Astronomia i Meteorologia (IEEC-UB), Institut Ci\`encies Cosmos, U. Barcelona, Mart\'i i Franqu\`es 1, 08028 Barcelona, Spain\\
  $^{4}$ Space Telescope Science Institute, 3700 San Martin Drive, Baltimore, MD 21218, USA\\
  $^{5}$ CSIRO Astronomy and Space Science, PO Box 76, NSW 1710 Epping, Australia\\
  $^{6}$ Departamento de Astronom\'ia, Universidad de Chile, Camino El Observatorio 1515, Las Condes, Santiago, Chile\\
  $^{7}$ Instituto de Astrof\'isica de Andaluc\'ia, CSIC, Glorieta de la Astronom\'ia, s/n, E-18008 Granada,Spain\\
  $^{8}$ INAF - Osservatorio Astrofisico di Arcetri, L.go E. Fermi 5, 50125, Firenze, Italy}
\begin{document}

\date{Accepted date. Received date; in original form date}


\maketitle


\begin{abstract}
We combine previously published interferometric and single-dish data of relatively nearby massive dense cores that are actively forming stars to test whether their `fragmentation level' is controlled by turbulent or thermal support. We find no clear correlation between the fragmentation level and velocity dispersion, nor between the observed number of fragments and the number of fragments expected when the gravitationally unstable mass is calculated including various prescriptions for `turbulent support'.
On the other hand, the best correlation is found for the case of pure thermal Jeans fragmentation, for which we infer a core formation efficiency around 13\%, consistent with previous works. We conclude that the dominant factor determining the fragmentation level of star-forming massive dense cores at 0.1~pc scale seems to be thermal Jeans fragmentation.
\end{abstract}


\section{Introduction}\label{sec:intro}

For more than 60 years it has been thought that turbulence is an agent capable of providing support to molecular clouds against gravitational collapse (\eg\ Chandrasekhar 1951, Bonazzola et al. 1987, L\'eorat et al. 1990, McKee \& Tan 2003), while simultaneously producing local density enhancements that may become Jeans-unstable and collapse (\eg\ Sasao 1973, Elmegreen 1993, Padoan 1995, V\'azquez-Semadeni \& Gazol 1995, Klessen et al. 2000, V{\'a}zquez-Semadeni et al. 2003), and this is currently the most accepted scenario for the dynamical state of molecular clouds (\eg\ Mac Low \& Klessen 2004; Krumholz \& McKee 2005;  Hennebelle \& Chabrier 2008, 2011; Hopkins 2012; Chabrier et al. 2014; Federrath 2015; Guszejnov \& Hopkins 2015; Salim, Federrath, \& Kewley 2015).
However, some recent studies suggest that this may not be the case.  
For instance, molecular clouds seem to form by large-scale compressions in the diffuse, warm, HI medium. The compressed gas undergoes a transition to the cold, dense atomic phase (\eg\ Hennebelle \& Perault 1999, Heitsch et al. 2005, V{\'a}zquez-Semadeni et al. 2006), which is highly prone to Jeans instability (Hartmann et al. 2001), and thus must begin soon to collapse, in spite of the turbulence generated inside it by the original compression (Koyama \& Inutsuka 2002, Audit \& Hennebelle 2005, V{\'a}zquez-Semadeni et al. 2007, Heitsch \& Hartmann 2008). Moreover, once molecular clouds achieve column densities $\sim 10^{21}$ cm$^{-2}$, they are able to form molecular gas (Bergin et al. 2004) so that the formation of molecules may be essentially a byproduct of the gravitational collapse of the clouds (Hartmann et al. 2001). In this alternative scenario, the observed non-thermal motions of molecular clouds,
rather than consisting of random, small-scale, isotropic motions that can act as a pressure, would actually be dominated by inward motions {\it caused} by the gravitational collapse, which would occur both at large and small scales in a hierarchical and chaotic fashion (V{\'a}zquez-Semadeni et al. 2009, Ballesteros-Paredes et al. 2011a). This implies that the bulk of the observed non-thermal motions cannot provide support against the self-gravity of the clouds.

In previous papers of this series, we have presented evidence that the dynamics of molecular clouds are indeed dominated by gravity by showing that this scenario unifies molecular clouds and massive clumps in a single scaling relation (Heyer et al. 2009) that extends those by Larson (1981,  Ballesteros-Paredes et al. 2011a), and by showing that numerical simulations of cloud evolution including self-gravity develop power-law high-density tails in their column density probability distribution functions as a consequence of the gravitational collapse (see also Klessen 2000, Kritsuk et al.\ 2011, Ballesteros-Paredes et al. 2011b; Federrath \& Klessen 2013), in agreement with observations (\eg\ Kainulainen et al. 2009; Schneider et al. 2013). In the present contribution, we present a further line of evidence, by examining the mechanism responsible for fragmentation of dense cores. Indeed, a still unsolved and highly debated question is what are the main drivers of fragmentation in massive dense cores\footnote{We will follow the nomenclature of Williams et al. (2000) and Bontemps et al. (2010) where a massive dense core refers to a dense gas structure of $\sim0.1$ pc in size and $\ga20$~\mo\ in mass, which does not necessarily collapse into one star but can fragment into compact condensations and form a small cluster of stars.}, which are believed to be the precursors of stellar clusters.
The crucial parameter to estimate the fragmentation level of a dense core is the Jeans mass, which in its general form takes into account the different mechanisms of support against gravity, through the use of the `effective sound speed', $c_\mathrm{eff}$ (\eg\ Mac Low \& Klessen 2004).
Among the most debated forms of support are turbulent and thermal support. Thus, if the average turbulence level, average temperature and average density of a massive dense core are known, one can easily calculate the Jeans mass in both cases and estimate, given the mass of the core, the number of fragments expected in each case, so that one can assess which form of support against gravity is controlling the fragmentation process. 

Up to now, this simple question could not be answered because of a lack of statistically significant samples of massive dense cores where the fragmentation level has been assessed in a uniform way and down to spatial resolutions comparable to separations between cluster members ($\sim1000$~AU).  Recently,  Palau et al. (2013, 2014) compiled a sample of 19 massive dense cores with on-going star formation and studied the fragmentation level within the cores down to $\sim1000$~au, 
and $\sim0.5$~\mo\ of mass sensitivity. This would be a first approach to study the number of protostars (compact fragments will most likely become protostars, see Palau et al. 2013, 2014) within a massive dense core (assumed to become a cluster). In these works, the fragmentation level was assessed by counting the number of millimetre sources within a field of view of 0.1~pc of diameter\footnote{The size of 0.1~pc was taken to be the smallest size where fragmentation could be studied (because of the limitations by the primary beam response) in the works of Palau et al. (2013, 2014).}, \Nmm. 
Furthermore, by fitting the spectral energy distributions and submillimetre intensity profiles of the cores, Palau et al. (2014) modelled their density and temperature structure, so that densities and temperatures at different spatial scales could be estimated and thermal Jeans masses could be calculated. In this work, we compile observational data based on dense gas tracers for the sample of cores of Palau et al. (2013, 2014) and analyze them in a uniform way, in order to assess the turbulence level, estimate the turbulent Jeans mass, and finally compare the fragmentation level observed to the fragmentation level expected for each form of support, turbulent or thermal.


The plan of the present contribution is as follows: \S\ref{sobs}
presents the compiled data we used. In \S\ref{sec:results} we present
the fragmentation level of our cores, and how this fragmentation correlates (or not) with their
physical parameters.  {Finally, i}n \S\ref{sec:discussion} we discuss the physical implications of our results and present our main conclusions.

\begin{table*}
\caption{Modelled properties of the massive dense cores and compiled observational velocity dispersions}
\begin{center}
{\small
\begin{tabular}{lccc cccc ccc}
\noalign{\smallskip}
\hline\noalign{\smallskip}
&
&$M_\mathrm{0.1pc}$\supa
&$n_\mathrm{0.1pc}$\supa
&$T_\mathrm{0.1pc}$\supa
&$\sigma^\mathrm{NH_3}_\mathrm{1D,obs}$\supb
&$\sigma^\mathrm{NH_3}_\mathrm{1D,nth}$\supb
&
&$\sigma^\mathrm{N_2H^+}_\mathrm{1D,obs}$\supb
&$\sigma^\mathrm{N_2H^+}_\mathrm{1D,nth}$\supb
\\
Source
&\Nmm\
&(\mo)
&(10$^5$\,cm$^{-3}$)
&(K)
&(\kms)
&(\kms)
&$\mathcal{M}_\mathrm{NH_3}$\supb
&(\kms)
&(\kms)
&$\mathcal{M}_\mathrm{N_2H^+}$\supb
\\
\noalign{\smallskip}
\hline\noalign{\smallskip}
1-IC1396N       		&4	&11	&3.6		&25	&$-$		&$-$		&$-$	&0.79	&0.78	&4.5	\\
2-I22198\supc        	&1.5	&11	&3.6		&26	&0.47	&0.45	&2.6	&0.59	&0.59	&3.4\\
3-N2071-1		&4	&17	&5.7	   	&24	&0.44	&0.43	&2.5	&$-$		&$-$		&$-$\\
4-N7129-2		&1	&11	&3.6	    	&35	&$-$		&$-$		&$-$	&0.59	&0.58	&2.8\\
5-CB3-mm		&2	&15	&5.2		&40	&$-$		&$-$		&$-$	&0.73	&0.72	&3.3\\
6-I22172N      		&3	&9	&3.2      	&48	&0.59	&0.58	&2.4	&0.87	&0.86	&3.6\\
7-OMC-1S      		&9	&38	&13 		&49	&1.11	&1.10	&4.5	&0.90	&0.89	&3.7\\
8-A5142         		&7	&39	&13	  	&47	&1.61	&1.61	&6.8	&1.09	&1.08	&4.6\\
9-I05358NE		&4	&27	&9.1	     	&35	&0.72	&0.71	&3.5	&1.07	&1.07	&5.3\\
10-I20126			&1	&14	&4.8	    	&68	&2.00	&1.99	&7.0	&0.85	&0.84	&2.9\\
11-I22134      		&3.5	&10	&3.2	       	&50	&0.71	&0.70	&2.8	&0.62	&0.61	&2.5\\ 
12-HH80-81         	&3 	&12	&4.2	      	&66	&0.74	&0.72	&2.6	&$-$		&$-$		&$-$\\ 	
13-W3IRS5          	&3.5	&12	&4.0		&138&0.87	&0.84	&2.1	&1.18	&1.17	&2.9\\
14-A2591			&1.5	&16 &5.2		&147&0.68	&0.62	&1.5	&$-$		&$-$		&$-$\\
\hline	
15-Cyg-N53		&6	&30	&10		&27	&0.17\supd&0.13\supd&0.7\supd&0.81	&0.80&4.4\\
16-Cyg-N12		&2.5	&15 	&5.0		&29	&$-$		&$-$		&$-$	&1.23	&1.23	&6.6\\
17-Cyg-N63		&1	&14	&4.6		&31	&$-$		&$-$		&$-$	&0.82	&0.82	&4.2\\
18-Cyg-N48		&5	&35	&12		&36	&1.25	&1.25	&6.1	&1.21	&1.21	&5.9\\
19-DR21-OH		&11	&69	&23		&49	&1.51	&1.51	&6.3	&$-$		&$-$		&$-$\\
\hline
\end{tabular}
\begin{list}{}{}
\item[$^\mathrm{a}$] $M_\mathrm{0.1pc}$ is the mass inside a region of 0.1~pc of diameter computed according: 
$M_\mathrm{0.1pc} = M(R=0.05\mathrm{pc})= 4\pi\,\rho_0\,r_0^p\,\frac{R^{3-p}}{3-p}$, where $p$, $r_0$, and $\rho_0$ are index of the density power law, the reference radius adopted to be 1000~AU, and the density at the reference radius (given in Table 1 of Palau et al. 2014); 
$n_\mathrm{0.1pc}$ and $T_\mathrm{0.1pc}$ correspond to average density and temperature inside a region of 0.1~pc of diameter. $T_\mathrm{0.1~pc}$ is estimated as
$T_\mathrm{R} = \frac{\int_0^R T(r)\rho(r)r^2 dr}{\int_0^R \rho(r)r^2 dr}$, where $T(r)$ and $\rho(r)$ were calculated as power laws of the form $T(r)=T_\mathrm{0}(r/r_0)^{-q}$ and $\rho(r)=\rho_0(r/r_0)^{-p}$, with $T_0$ and $\rho_0$ being the values at the reference radius $r_0$ of 1000 AU. $T_0$, $\rho_0$, $p$, and $q$ are given in Table 1 of Palau et al. (2014). The final expression is
$T_\mathrm{R}=\frac{T_0(3-p)}{3-p-q}\left(\frac{r}{r_0}\right)^{-q}$.
\item[$^\mathrm{b}$]  $\sigma^\mathrm{NH_3}_\mathrm{1D,obs}$ and $\sigma^\mathrm{N2H^+}_\mathrm{1D,obs}$ are calculated from the measured FWHM line width, $\Delta v_\mathrm{obs}$,  as $\sigma_\mathrm{1D,obs}=\Delta v_\mathrm{obs}/(8\,\mathrm{ln}2)^{1/2}$. $\sigma_\mathrm{1D,nth} =\sqrt{\sigma_\mathrm{1D,obs}^2-\sigma_\mathrm{th}^2}$, with $\sigma_\mathrm{th}=\sqrt{k_\mathrm{B}\,T/(\mu\,m_\mathrm{H})}$ ($k_\mathrm{B}$ being the Boltzmann constant, $\mu$ the molecular weight (17 for \nh, 29 for \nth), $m_\mathrm{H}$ the mass of the hydrogen atom and $T$ the temperature of the region, taken from column (5) of this table).
The Mach number $\mathcal{M}$ is calculated as $\sigma_\mathrm{3D,nth}$/$c_\mathrm{s}$, with  $c_\mathrm{s}$ being the sound speed calculated as $c_\mathrm{s}=\sqrt{k_\mathrm{B}\,T/(\mu\,m_\mathrm{H})}$, using $\mu=2.3$, and $\sigma_\mathrm{3D,nth}=\sqrt{3}\,\sigma_\mathrm{1D,nth}$.
\item[$^\mathrm{c}$] The parameters of the density and temperature structure for this source are different from Palau et al. (2014) because here we have used the original JCMT data of Jenness et al. (1995) and we have not assumed any error beam in the modelling (see Appendix B). 
\item[$^\mathrm{d}$] Marginal detection of the \nh(1,1) line, not taken into account in the analysis of this work.
\item[] Refs: IC\,1396N: Alonso-Albi et al. (2010); 
I22198: S\'anchez-Monge et al. (2013), Fontani et al. (2011); 
NGC2071-1: Zhou et al. (1990);
NGC7129-2: Fuente et al. (2005);
CB3-mm: Alonso-Albi et al. (2010); 
I22172N: S\'anchez-Monge et al. (2013); Fontani et al. (2006);
OMC-1S: Wiseman \& Ho (1998); Tatematsu et al. (2008);
A5142: S\'anchez-Monge et al. (2013); Fontani et al. (2011);
I05358NE: S\'anchez-Monge et al. (2013); Fontani et al. (2011);
I20126: S\'anchez-Monge et al. (2013); Fontani et al. (2006);
I22134: S\'anchez-Monge et al. (2013); Fontani et al. (2015);
HH80-81: G\'omez et al. (2003); 
W3IRS5: Tieftrunk et al. (1998); Gerner et al. (2014);
A2591: Torrelles et al. (1989);
Cyg-N53: VLA archive; Bontemps et al. (2010);
Cyg-N12: Bontemps et al. (2010);
Cyg-N63: Bontemps et al. (2010);
Cyg-N48: Mangum et al. (1992), Bontemps et al. (2010);
DR21-OH: Mangum et al. (1992).
\end{list}
}
\end{center}
\label{tveldisp}
\end{table*}

\section{The sample and data compilation}\label{sobs}

The present work is based on the sample of massive dense cores presented in Palau et al. (2013, 2014), whose distances, luminosities, and masses range from 0.45 to 3~kpc, from 300 to 10$^5$~\lo, and from 80 to 1500~\mo, respectively (given in Table~A1). 
The sample was selected from deeply embedded intermediate/high-mass star-forming regions published in the literature that have been studied in the millimetre range down to mass sensitivities of $\sim0.5$~\mo, and spatial resolutions of $\sim1000$~au.
Palau et al. (2014) modelled the massive dense cores with
temperature and density profiles decreasing with radius following
power-laws, and determined a number of properties of the density and
temperature structure in a uniform way for all the sample. 

In order to estimate the turbulence level in each core in a uniform way and compare it
to the `turbulent fragmentation' level, we used Very Large Array (VLA) \nh(1,1)
data in C/D configuration, available for 14 (out of 19) massive dense
cores (Torrelles et al. 1989; Zhou et al. 1990; Mangum et al. 1992;
Tieftrunk et al. 1998; Wiseman \& Ho 1998; G\'omez et al. 2003;
S\'anchez-Monge et al. 2013; see also Palau et al. 2014). VLA beams
are typically $\la5$~arcsec, and the minimum angular scales filtered
out by the interferometer are $\ga35$~arcsec (estimated following
appendix in Palau et al. 2010, and using a minimum baseline of
35~m). {The latter} corresponds to $\sim0.3$~pc (for typical distances of the regions in the sample), slightly larger than
the field of view of 0.1~pc that we are studying. Thus, with these VLA
\nh\ data we are recovering most of the emission at the spatial scales
we are studying, with an angular resolution good enough ($\sim5$~arcsec) to resolve
typical sizes of massive dense cores (\eg\ S\'anchez-Monge et
al. 2013). 
The \nh(1,1) hyperfine structure was fitted and hyperfine
`observed' FWHM line widths, $\Delta v_\mathrm{obs}$, were inferred for each core. 
We note that we took special care to make the sample as uniform as possible and we measured the \nh(1,1) line width in all cases by using the `nh3(1,1) method' in the CLASS package of the GILDAS software, on the spectrum resulting from averaging the \nh(1,1) emission over the central region of $\sim0.1$~pc of diameter of the dense core where we study the fragmentation level. 
When literature did not provide an average spectrum over the massive dense core that we are studying, we downloaded and reduced the VLA data to extract the spectrum. This was done for HH\,80-81, W3\,IRS5, A2591, Cyg-N53, Cyg-N48, and DR21-OH (VLA projects AG0552, AT0180, AT0084, AW0240, AF386, respectively: standard calibration as described in S\'anchez-Monge et al. 2013 was applied).
In two sources (I22134 and DR21-OH) we fitted two velocity components.
From these line widths we calculated the observed velocity dispersions as
$\sigma_\mathrm{1D,obs}=\Delta
v_\mathrm{obs}/(8\,\mathrm{ln}2)^{1/2}$. The values of the compiled
observed velocity dispersions from \nh(1,1) VLA data are listed in
Table~\ref{tveldisp}, and the spectra with the corresponding fits to the hyperfine structure are shown in Appendix B.

Because in some cases the VLA \nh(1,1) emission might be affected by the passage of an outflow (\eg\
the \nh(1,1) line width of IRAS\,20126+4104 is larger along the
direction of the outflow, see Fig.~B1 of S\'anchez-Monge et al. 2013),
we additionally compiled data from a different dense gas tracer,
\nth(1--0), observed using a single-dish telescope (IRAM\,30m in all cases except for OMC-1S, for which Nobeyama\,45m was used). 
This is a reasonable approach to avoid contamination by outflow for several reasons. 
First, \nth\ is known to be destroyed by CO (\eg\ Joergensen 2004; Busquet et al. 2011). 
Second, the outflow is typically compact and thus its emission should be diluted in the single-dish beam. 
Thus, the \nth\ line widths should be less affected by the passage of the outflow and thus more reliable to measure the `initial' non-thermal motions unaffected by stellar feedback.
The \nth(1--0) data were available for 15 out of 19 regions and were compiled from the literature (Fuente et al. 2005;
Fontani et al. 2006, 2011, 2015; Tatematsu et al. 2008; Alonso-Albi et al. 2010; Bontemps et al. 2010; Gerner et al. 2014), and its hyperfine structure was fitted using the CLASS package of the GILDAS software. 
The IRAM\,30m Telescope provides a beam of $\sim26$~arcsec at the frequency of
\nth(1--0), comparable to the spatial scale at which the massive dense cores are being studied (0.1~pc, at
the typical distances of the cores of our sample), and about a factor of 5 larger than the VLA beam.  
By using the same method outlined above for \nh(1,1) we inferred the observed velocity
dispersions for \nth(1--0), listed in Table~\ref{tveldisp} and the spectra and fits are shown in Appendix B. 
The method used to fit the hyperfine structure for both \nth\ and \nh\ takes the opacity effects into account.
The observed \nth(1--0) velocity dispersions range from 0.6 to 1.2~\kms, a
narrow{er} range {than that of the} velocity dispersions inferred
from VLA \nh\ data (ranging from 0.5 to 2.0~\kms), as expected
(because of {the} outflow contamination of the \nh\ VLA data).


The main difference between the line widths reported in this work and the line widths reported in column (10) of Table~2 of Palau et al. (2014\footnote{Linewidths reported in column (9) of Table~2 of Palau et al. (2014), or in column (10) of Table~4 of Palau et al. (2013) correspond to quiescent cores in the surroundings of the massive dense cores where fragmentation is being studied, and are not comparable to the line widths reported here, corresponding to the massive dense cores where active star formation is taking place and where fragmentation is being studied.}) is that we here re-reduced the interferometric data and re-did the fits of the spectra in all cases using the same method, instead of just taking the values reported in the literature, which use different methods. Thus the present analysis is uniform in the sense that the method to infer the line widths is the same for all sources.

\begin{figure}
\begin{center}
\begin{tabular}[b]{c}
     \epsfig{file=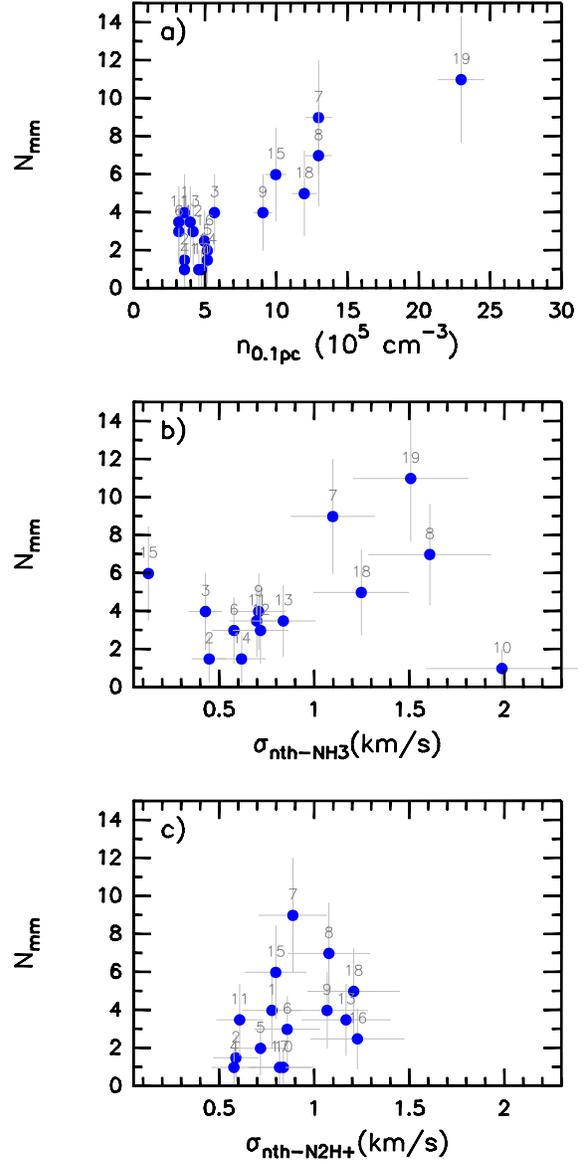, width=7.5cm,angle=0} \\
\end{tabular}
\caption{Observed `fragmentation level' ($N\mathrm{mm}$) vs different quantities (Table~\ref{tveldisp}). 
{\bf a)} \Nmm\ vs the density of the core within a region of 0.1~pc of diameter. 
{\bf b)} \Nmm\ vs the non-thermal velocity dispersion as inferred from VLA \nh(1,1) data. 
{\bf c)} \Nmm\ vs the non-thermal velocity dispersion as inferred from single-dish \nth\,(1--0) data. 
}   
\label{fNmmnsigma}
\end{center}
\end{figure}

\section{Results and analysis}\label{sec:results}

\subsection{Fragmentation level vs.\ density and velocity dispersions}\label{sec:res-frag}

The compiled observed velocity dispersions of \nh(1,1) and \nth(1--0)
together with the modeling of the temperature structure of the massive
dense cores (Palau et al. 2014) allowed us to separate the thermal
from the non-thermal contribution of the observed velocity dispersion.
We estimated the thermal component of the velocity dispersion,
$\sigma_\mathrm{th}$, from
$\sqrt{k_\mathrm{B}\,T/(\mu\,m_\mathrm{H})}$, with $k_\mathrm{B}$ the
Boltzmann constant, $\mu$ the molecular weight (17 for \nh, 29 for
\nth), $m_\mathrm{H}$ the mass of the hydrogen atom, and $T$ the
temperature of the region, which was adopted from the average
density-weighted temperature inside a region of 0.1~pc of
diameter (the same region where we assessed the fragmentation level). 
This average temperature is estimated from the density
and temperature power-laws modelled by Palau et al. (2014, see notes of
Table~\ref{tveldisp} for further details) for each core.
The non-thermal component was estimated by using $\sigma_\mathrm{1D,nth}
=\sqrt{\sigma_\mathrm{obs}^2-\sigma_\mathrm{th}^2}$.  
Then, the total (thermal + non-thermal) velocity dispersion is calculated by adding quadratically the thermal and non-thermal components, using for the thermal component a molecular weight of 2.3, which corresponds to the sound speed and thus:
$\sigma_\mathrm{1D,tot} =\sqrt{c_\mathrm{s}^2+\sigma_\mathrm{nth}^2}$.
%
The Mach number $\mathcal{M}$ is calculated as
$\sigma_\mathrm{3D,nth}/c_\mathrm{s}$, with $\sigma_\mathrm{3D,nth}=\sqrt{3}\,\sigma_\mathrm{1D,nth}$. 
The resulting Mach numbers range from $\sim2$ to 7.

In Fig.~\ref{fNmmnsigma} we plot the number of millimetre sources
within a field of view of 0.1~pc in diameter, \Nmm\ (a proxy to the
fragmentation level), as function of (a) the core density within a
region of 0.1~pc of diameter (modelled in Palau et al. 2014) ---upper
panel---, (b) the non-thermal velocity dispersion
measured from \nh(1,1) (interferometric data) ---middle panel---, and {from}
\nth(1--0) (single-dish data) ---lower panel---. The figure shows that,
while there is a correlation between \Nmm\ and the density within
0.1~pc (correlation coefficient: 0.89), there is no clear trend between \Nmm\ and the velocity
dispersion measured with \nh\ (correlation coefficient: 0.27), nor with \nth\ (correlation coefficient: 0.35). 
This suggests that the velocity dispersion of the massive dense cores
might not be a crucial ingredient in determining the fragmentation level.

\begin{figure*}
\begin{center}
\begin{tabular}[b]{c}
     \epsfig{file=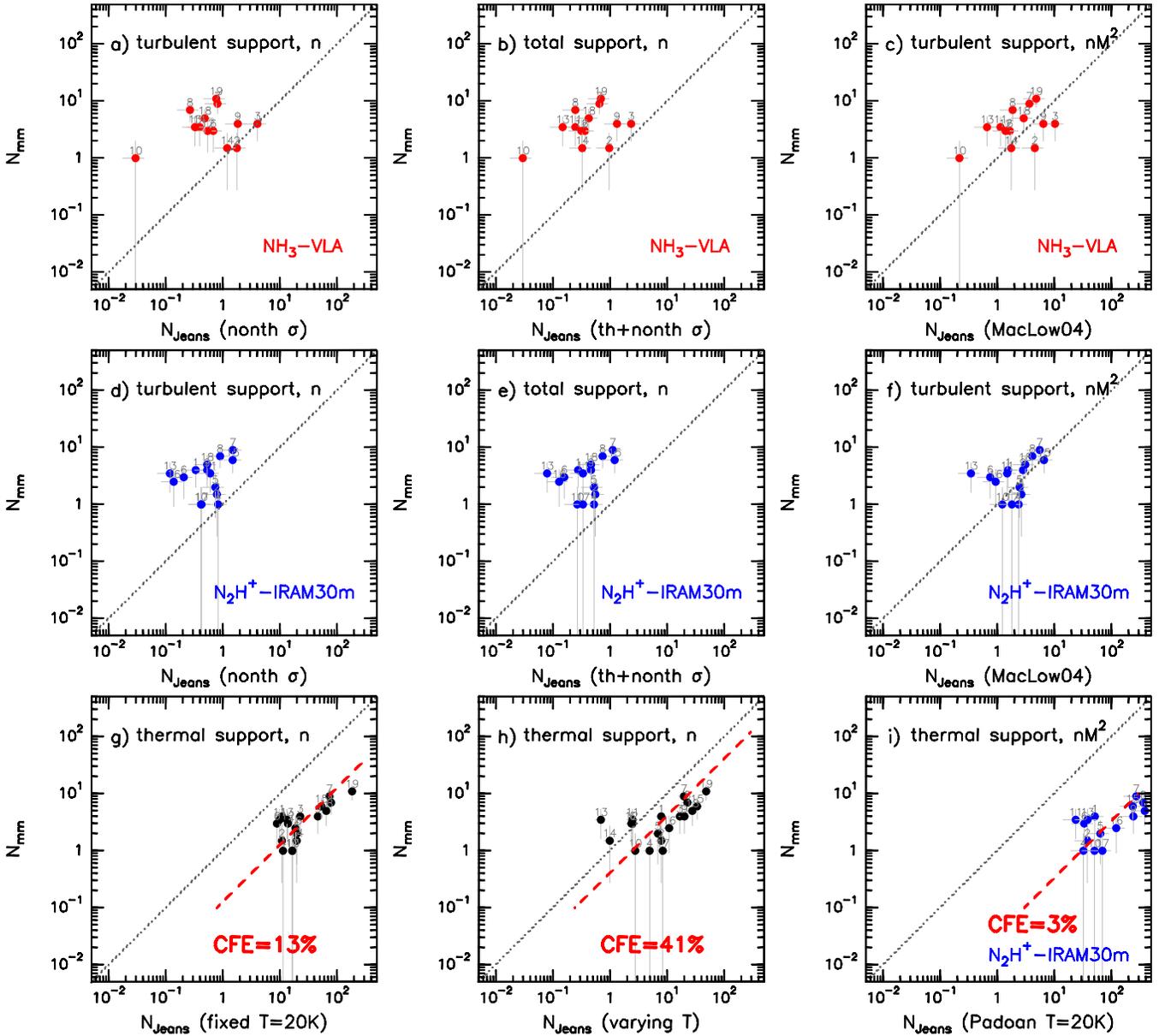, width=16cm,angle=270} \\
\end{tabular}
\caption{`Fragmentation level' ($N\mathrm{mm}$) vs Jeans number.  
{\bf Top panels:} Jeans number calculated using the velocity dispersions estimated from \nh, as explained in Section~3.2.  
{\bf Middle panels:} idem but using the velocity dispersions estimated from \nth\ (Section~3.2).  
{\bf Bottom panels:} Jeans number calculated considering only thermal support, calculated either using a fixed temperature of 20~K for all the cores (panels `g' and `i'), or the temperature inferred from the core modelling presented in Palau et al. (2014; panel `h'; see Section 3.3). 
In all panels, the Jeans number is calculated using the average density inside a region of 0.1~pc of diameter (as explained in the table notes of Table~1;
see also Palau et al. 2014), except for panels on the right, where the average density has been multiplied by the square of the Mach number (following Mac Low \& Klessen 2004).
In all panels, the dotted black line represents the one-to-one relation, for a core formation efficiency (CFE) of unity. For `g', `h', and `i' panels, the red dashed line corresponds to the fit with slope$=1$ used to infer the indicated CFE (3--41\%).
}
\label{fNjeans}
\end{center}
\end{figure*}

\begin{table*}
\caption{Linear fits to the \Nmm\ vs \Nj\ relations of Fig.~\ref{fNjeans}, which correspond to different cases of core support}
\begin{center}
{\small
\begin{tabular}{lcccccc}
\noalign{\smallskip}
\hline\noalign{\smallskip}
&
&panels
&correlation
&
&CFE \supc
&
\\
Support \supa
&equation \supa
&Fig.~2 \supa
&coefficient \supb
&slope \supb
&(\%)
&$\chi^2$ \supc
\\
\noalign{\smallskip}
\hline\noalign{\smallskip}
turbulent (\nh) 		     					&(2)	&a, d	&0.24	&$0.14\pm0.18$	&$>100\%$	&$-$		\\
turbulent (\nth)							&(2)	&a, d	&0.23	&$0.22\pm0.25$	&$>100\%$	&$-$		\\
turbulent $n\mathcal{M}^2$ (\nh) 			&(5)	&c, f	&0.50	&$0.34\pm0.18$	&$\ga100\%$	&$-$		\\
turbulent $n\mathcal{M}^2$ (\nth)  			&(5)	&c, f	&0.36	&$0.33\pm0.24$	&$\ga100\%$	&$-$		\\
thermal T=20K     						&(6)	&g	&0.72	&$0.60\pm0.14$	&$13\%$		&1.27	\\
thermal varying T		     				&(6)	&h	&0.57	&$0.34\pm0.12$	&$41\%$		&3.34	\\
thermal T=20K  $n\mathcal{M}^2$ (\nh)		&(8)	&$-$	&0.69	&$0.35\pm0.11$	&$3.7\%$		&2.37	\\
thermal T=20K  $n\mathcal{M}^2$ (\nth) 		&(8)	&i	&0.64	&$0.47\pm0.15$	&$3.3\%$		&1.49	\\
\hline
\end{tabular}
\begin{list}{}{}
\item[$^\mathrm{a}$] Fits are performed for the relations of the panels (of Fig.~2) indicated in column (3), which correspond to \Nj\ estimated using the equations given in column (2).
\item[$^\mathrm{b}$] Correlation coefficient and slope of a linear fit with two free parameters. The slope should be close to one if the form of support correctly described the observations.
\item[$^\mathrm{c}$] Core Formation Efficiency inferred forcing a linear fit with slope = 1, and the corresponding $\chi^2$.
\end{list}
}
\end{center}
\label{tNmmNj}
\end{table*}

\subsection{Fragmentation level vs `turbulent' Jeans number}\label{sec:fragm-turb}

To further compare the role of the physical properties of the cores (density, temperature, velocity dispersion) in determining the
fragmentation, we estimated the expected number of fragments under different assumptions for the gravitationally unstable mass, to which, for convenience, we continue referring as a `Jeans' mass in general.  
To estimate the Jeans mass, we started from the Jeans length, $L_\mathrm{Jeans}= \sqrt{\frac{\pi c_\mathrm{eff}^2}{G\rho_\mathrm{eff}}}$ (\eg\ Kippenhahn, Weigert \& Weiss 2012), and assumed spherical symmetry, $M_\mathrm{Jeans}=  \frac{4\pi}{3} \rho_\mathrm{eff}  \big(\frac{L_\mathrm{Jeans}}{2}\big)^3$, where $c_\mathrm{eff}$ is the `effective sound speed', $\rho_\mathrm{eff}$ is the `effective density', and $G$ is the gravitational constant. Therefore:

\begin{equation}
M_\mathrm{Jeans}=  \frac{\pi^{5/2}}{6\,G^{3/2}}\, c_\mathrm{eff}^{3}\,\rho_\mathrm{eff}^{-1/2}.
\end{equation}

First, we have searched for a correlation between the observed fragmentation level \Nmm\ and the expected number of fragments in a
turbulent support scenario (\eg\ Mac Low \& Klessen 2004). Thus, we have computed the expected mass of the fragments by assuming that the critical `Jeans' mass is determined by non-thermal (`turbulent') support, where the `effective sound speed' $c_\mathrm{eff}$ corresponds to the non-thermal component of the observed velocity dispersion $\sigma_\mathrm{1D,nth}$. This, in practical units and using the number density of H$_2$ molecules (as calculated by the model in Palau et al. 2014, and using a molecular weight of 2.8), can be written as:

\begin{equation}
\Big[\frac{M_\mathrm{Jeans}^\mathrm{nth}}{M_{\sun}}\Big] = 
0.8255\,\Big[\frac{\sigma_\mathrm{1D,nth}}{0.188\,\mathrm{km\,s}^{-1}}\Big]^3
\Big[\frac{n_\mathrm{H_2}}{10^5\,\mathrm{cm}^{-3}}\Big]^{-1/2}.
\end{equation}

Then, the number of expected fragments, \Nj, is estimated from the ratio of the mass of the core inside a region of 0.1~pc of diameter, $M_\mathrm{0.1pc}$\footnote{The mass inside a region of 0.1~pc of diameter is typically $\sim10$\% of the total mass of the core (given in Table~4 of Palau et al. 2013) and is only marginally correlated to the mass of the core, in part because far from the central part of the core, the core departs from sphericity, a basic assumption of the core modelling of Palau et al. (2014).}, and the Jeans mass,
$M_\mathrm{Jeans}$:

\begin{equation}
N_\mathrm{Jeans} = 
\frac{M_\mathrm{0.1pc}}{M_\mathrm{Jeans}}.
\end{equation}

The result is presented in Figs.~\ref{fNjeans}a and \ref{fNjeans}d (for \nh(1,1) and \nth\,(1--0), respectively). 

We also estimated the Jeans mass including both thermal and non-thermal support:

\begin{equation}
\Big[\frac{M_\mathrm{Jeans}^\mathrm{tot}}{M_{\sun}}\Big] = 
0.8255\,\Big[\frac{\sigma_\mathrm{1D,tot}}{0.188\,\mathrm{km\,s}^{-1}}\Big]^3
\Big[\frac{n_\mathrm{H_2}}{10^5\,\mathrm{cm}^{-3}}\Big]^{-1/2}
\end{equation}
(Figs.~\ref{fNjeans}b and \ref{fNjeans}e), and including only non-thermal support but taking into account that large-scale supersonic flows compress the gas and generate density enhancements, which are the ones that proceed to collapse (Mac Low \& Klessen 2004). In this case, the `effective density' is obtained by multiplying the average density of the core by the square of the Mach number:

\begin{equation}
\Big[\frac{M_\mathrm{Jeans}^\mathrm{conv.flows}}{M_{\sun}}\Big] = 
0.8255\,\Big[\frac{\sigma_\mathrm{1D,nth}}{0.188\,\mathrm{km\,s}^{-1}}\Big]^3
\Big[\frac{n_\mathrm{H_2}\,\mathcal{M}^2}{10^5\,\mathrm{cm}^{-3}}\Big]^{-1/2},
\end{equation}
(Figs.~\ref{fNjeans}c and \ref{fNjeans}f).

Figs.~\ref{fNjeans}a (\nh) and \ref{fNjeans}d (\nth) reveal a very weak correlation of \Nmm\ with \Nj\ (correlation coefficient of 0.24 and 0.23, respectively), and a slope $\ll1$ ($0.14\pm0.18$ for \nh, and $0.22\pm0.25$ for \nth, see Table~\ref{tNmmNj}). For the case of non-thermal+thermal support the situation is very similar (Figs.~\ref{fNjeans}b and \ref{fNjeans}e), and for the case of `density enhanced by turbulence' (equation (5) and Figs.~\ref{fNjeans}c and \ref{fNjeans}f), the correlation coefficient increases up to 0.50 and the slope up to $0.34\pm0.18$  (Table~\ref{tNmmNj}), with respect to the case of turbulence providing only support (equations (2) and (4)). This was expected, because when taking into account the density enhancements produced by turbulence the role of turbulence providing support becomes less important and the correlation slightly improves. 

Therefore, we show that considering the non-thermal motions as a form of support does not provide a good correlation between the expected number of fragments and the observed number in any of the cases considered.
More importantly, we note that the turbulent Jeans number for the majority of the cores is less than or similar to unity in all three cases, which would imply that, if turbulent support were active, these cores should not fragment at all, contrary to what is observed. 
As we will see in the next Section, this would imply a Core Formation Efficiency (CFE, see below) $\ga 100$\%, which is meaningless.

\subsection{Fragmentation level vs thermal Jeans number}\label{sec:fragm-th}

Given the poor correlations found between the observed number of fragments and the expected number of fragments in case of turbulent support, we considered only thermal support (no contribution from `turbulence'). In this case, the `effective sound speed' $c_\mathrm{eff}$ simply corresponds to the sound speed of the gas, which can be written in terms of the kinetic temperature, and the Jeans mass is finally written as:

\begin{equation}
\Big[\frac{M_\mathrm{Jeans}^\mathrm{th}}{M_{\sun}}\Big] = 
0.6285\,\Big[\frac{T}{10\,\mathrm{K}}\Big]^{3/2}
\Big[\frac{n_\mathrm{H_2}}{10^5\,\mathrm{cm}^{-3}}\Big]^{-1/2}.
\end{equation}

This was done using two different assumptions for the temperature. First, we used a fixed temperature of $\sim20$~K, as a first approximation to the `initial' (i.e., before being heated by the protostellar feedback) temperature of the dense core (\eg\ S\'anchez-Monge et al. 2013). Second, we used the average temperature estimated for each core within a region of 0.1~pc of diameter, $T_\mathrm{0.1pc}$ (ranging from 25 to 150~K, Table~\ref{tveldisp}). This assumption should give an upper limit to the temperature at the time when fragmentation took place\footnote{The stellar feedback should affect the density structure on larger timescales compared to the timescale when stellar feedback modifies the temperature because the first should change through mechanical processes while the latter changes through radiative processes. In addition, the massive dense cores of our sample are in similar evolutionary stages, having not developed UCHII regions yet (see Palau et al. 2014 for a more detailed discussion), and the Jeans mass depends more strongly on temperature than on density. For these reasons we consider that the density structure of the massive dense cores in our sample is a reasonable approach to the density structure at the time of fragmentation.}.

The results are plotted in Fig.~\ref{fNjeans}g,h. Fig.~\ref{fNjeans}g shows a correlation of \Nmm\ and \Nj, with a slope of $0.60\pm0.14$, clearly larger and closer to 1 than the slope obtained for the turbulence-supported case (0.2--0.3, Table~2). In this panel, the temperature is fixed for all the cores and equal to 20~K. The data are clearly offset with respect to the one-to-one relation (dotted black line), which can be explained if only a percentage of the total mass of the core is converted into compact fragments. We define the Core Formation Efficiency (CFE) as the fraction of mass of a dense core found in (pre- and proto-stellar) compact fragments (as in Bontemps et al. 2010), and in this case:
\begin{equation}
N_\mathrm{Jeans} = 
\frac{M_\mathrm{0.1pc}\,\mathrm{CFE}}{M_\mathrm{Jeans}}.
\label{CFE}
\end{equation}
Thus, we fitted a line with slope 1 {(dashed red line in Fig.~\ref{fNjeans})} and the offset should be a first approximation to the CFE.
By doing this for the dataset of Fig.~\ref{fNjeans}g we found a CFE of 13\%, with a correlation coefficient of 0.72.
This value for the CFE is fully consistent with the independent direct measurements
of the CFE by Bontemps et al. (2010) and Palau et al. (2013), who
estimated this quantity by dividing the total mass in compact fragments (detected with an
interferometer in an extended configuration) by the mass of the core (measured with a single-dish). Our inferred CFE is also similar to those found by Louvet et al. (2014) in the W43-MM1 region.

We additionally estimated \Nj\ using the different average temperatures inferred for each core (Table~\ref{tveldisp}), and the
result is shown in Fig.~\ref{fNjeans}h. In this approach,
\Nj\ is smaller (compared to the previous case of fixed temperature
equal to 20~K), because the Jeans masses are larger due to the higher
adopted temperatures, and hence the inferred CFE is larger as well. 
%
The effect of using these higher temperatures is to predict too small
{a} number of fragments (too small \Nj), especially for the two extreme cases (cores 13 and 14) which are also the most luminous regions.
In this case we obtained a correlation coefficient of 0.57, and a CFE of 41\%.

Finally, we calculated the Jeans mass considering that turbulence is only producing regions of higher density, but not providing support against gravity (with the latter being only thermal, \eg\ Padoan \& Nordlund 2002):

\begin{equation}
\Big[\frac{M_\mathrm{Jeans}^\mathrm{conv.flows-th}}{M_{\sun}}\Big] = 
0.6285\,\Big[\frac{T}{10\,\mathrm{K}}\Big]^{3/2}
\Big[\frac{n_\mathrm{H_2}\,\mathcal{M}^2}{10^5\,\mathrm{cm}^{-3}}\Big]^{-1/2}
\end{equation}
We studied this case using Mach numbers calculated from \nh\ and \nth\ data (Table~1), and the results are listed in Table~2. The fit performed using \nh\ (to estimate the Mach number) has a correlation coefficient very similar to the coefficient obtained for pure thermal support at a fixed temperature of 20~K in Fig.~2g, but the slope of the linear fit is significantly smaller ($0.35\pm0.11$) and thus deviates more strongly from the one-to-one relation. As for the fit performed using \nth\ (to estimate the Mach number), the slope is more similar to the slope in Fig.~2g. In Fig.~2i we show the case of \nth\ only for clarity. We also performed a linear fit forcing the slope to 1 to infer the CFE in this case, which is around 3\% for both \nh\ and \nth\ (see Table~2 and Fig.~2i). The CFE is very low because the densities in this case are higher (by $\mathcal{M}^2$) and the Jeans mass decreases resulting in a very high number of expected fragments.

Overall, the best correlation between \Nmm\ and \Nj\ is found for the case of pure thermal support adopting a temperature of $\sim20$~K for all the cores and with no modification of the density by the Mach number (Fig.~2g). In addition, also for this case the slope in the  \Nmm\ vs \Nj\ relation is closest to 1 (see Table~\ref{tNmmNj}).

\begin{figure}
\begin{center}
\begin{tabular}[b]{c}
     \epsfig{file=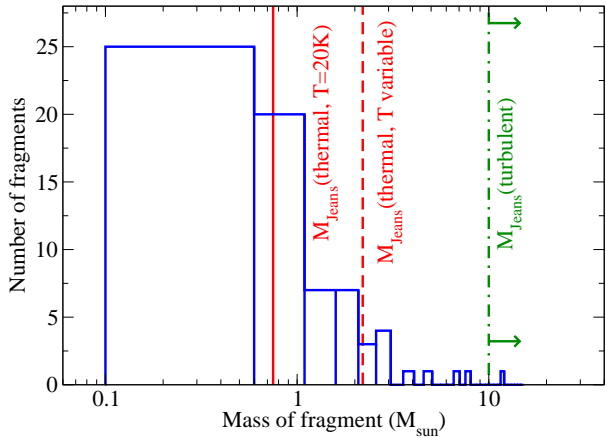, width=7.2cm,angle=270} \\
\end{tabular}
\caption{Histogram of fragments masses. The red solid line corresponds to the Jeans mass, averaged over all the sample, assuming thermal support at a fixed temperature of 20~K; the red dashed line corresponds to the Jeans mass, averaged over all the sample, assuming thermal support at the average temperature measured for each core inside a region of 0.1~pc of diameter (which should be an upper limit to the temperature at the time of fragmentation); and the green dotted-dashed line corresponds to the Jeans mass, averaged over all the sample, assuming turbulent support as described in equation (5) (Mac Low \& Klessen 2004). For the other prescriptions of turbulent support (Section~\ref{sec:fragm-turb}), we obtain Jeans masses up to $\sim80$~\mo.}   
\label{ffragmass}
\end{center}
\end{figure}

\subsection{Fragment masses}\label{sec:fragm-mass}

We have estimated the masses of the fragments identified in each massive dense core by assuming the temperature at the distance of the fragment (from the core center\footnote{The core center is taken as the peak of the millimetre/submillimetre emission observed with a single-dish (see Palau et al. 2014 for further details).}) as provided by our modelled envelopes. We used the dust opacity law of Ossenkopf \& Henning (1994, icy mantles for densities $\sim10^6$~\cmt). The results are shown in Fig.~\ref{ffragmass}. About 45 fragments (out of 75) present masses $<1$~\mo. Although these masses have been inferred from interferometric observations using extended configurations, and thus part of the flux must have been filtered out by the interferometer, we estimate that the missed flux is probably not larger than a factor of $\sim2$ (the reason for this is that any fragment detected by the interferometer at such extended configurations must be intrinsically compact). For example, for the case of I22198, S\'anchez-Monge et al. (2010) report a flux density at 1.3~mm using the SMA in compact configuration of $\sim500$~mJy, while Palau et al. (2013) report a flux density at 1.3~mm using the PdBI in its most extended configuration of $\sim270$~mJy (adding MM2 and MM2S). Thus, as a first reasonable approach, one might conclude that most of the fragments in our sample must have masses of the order of 1--2~\mo, or below.

On the other hand, the Jeans masses calculated for each region considering only thermal support for a fixed temperature of 20~K (shown as a solid line in Fig.~\ref{ffragmass}) range from  0.4 to 1~\mo, with an average value of 0.75~\mo. If we use our second approach for the temperature, \ie\ use the average temperature estimated for each core inside a region of 0.1~pc of diameter (Table~\ref{tveldisp}, without including cores 13 and 14, whose temperature is clearly affected by stellar feedback), we find Jeans masses in the range from 1 to 5~\mo, with an average Jeans mass for all the regions of 2.2~\mo\ (shown as a  dashed line in Fig.~\ref{ffragmass}). Since this last case should yield an upper limit to the temperature at the time of fragmentation, the Jeans mass in this case is also an upper limit. For the case of average Jeans masses including the turbulence as a support term (equations 2, 4, and 5), these range from 10~\mo\ (equation 5) up to $\sim80$~\mo\ (equations 2 and 4). Therefore, we conclude that the typical masses of most of the fragments, around 1~\mo, are in good agreement with the Jeans mass considering pure thermal support only, with no need of additional forms of support.

\section{Discussion and conclusions}\label{sec:discussion}

In the present work we have investigated which process, turbulence or gravity, is controlling the fragmentation level of massive dense cores at a scale of $\sim0.1$~pc. We have used a sample of 19 massive cores, previously presented in Palau et al. (2013, 2014), to show that the fragmentation of  these objects seems to be controlled mainly by thermal Jeans fragmentation. Specifically, we have shown that the fragmentation level, \Nmm\  ---measured as the number of compact fragments within a core---, presents a significantly better correlation with the core density than with its non-thermal velocity dispersion. Correspondingly, we have shown that \Nmm\ correlates linearly with the number of fragments expected from simple thermal Jeans fragmentation, with a core formation efficiency, CFE, around 13\%, while, instead, assuming turbulence-dominated fragmentation, the majority of the cores should not fragment, contrary to what is observed. Finally, we have given hints that the masses of most of the fragments in our sample seem to be of the order of the Jeans mass calculated considering only pure thermal support.
  
\subsection{Comparison to previous works}

It is important to point out that our results are not inconsistent with those of Zhang et al. (2009), Pillai et al. (2011), Wang et al. (2011, 2014), and Lu et al. (2015). 
Those authors concluded that the fragments within massive cores of infrared-dark clouds have masses significantly larger than the thermal Jeans mass, and consistent with the turbulent Jeans mass instead. However, in most of these clouds, this could be related to a sensitivity and spatial resolution issue due to the large distance of these infrared-dark clouds (ranging from 3.3 to 7.4 kpc). For example, in four (out of six) of those clouds, the mass sensitivity is  $>2$~\mo\ (above the Jeans mass) and the spatial resolution is $>5000$~au, while the massive dense cores studied here are all observed with mass sensitivities $<1$~\mo, and spatial resolutions $\sim1000$~au.

The most puzzling clouds are G28.34+0.06\,P1 (Zhang et al. 2015), and the Snake (G11.11$-$0.12\,P6, Wang et al. 2014). In these two clouds, observed down to subsolar mass sensitivities, there seems to be a lack of low-mass fragments, suggesting that the bulk of low-mass fragments have not formed yet at such earlier stages, or that the low-mass fragments form outside the core and follow the global collapse of the cloud. However, other recent multiwavelength studies focused on the stellar content of infrared-dark clouds show that most of the protostars formed in these clouds are of low-mass ($<2$~\mo; \eg\ Samal et al. 2015), and thus this needs to be further investigated.
Our study, carried out toward a sample of 19 regions more evolved than those in infrared-dark clouds, shows that most of the fragments detected in our sample are of low mass, and consistent with the thermal Jeans mass, indicating that at these stages the low-mass fragments do already exist. If the lack of low-mass fragments in infrared-dark clouds is confirmed in future observations, our data suggest that the duration of the stage when the low-mass fragments are formed is quite short. This is consistent with the extremely non-linear nature of the gravitational collapse (see, \eg\ Fig.~1 (bottom-left) of Toal\'a, V\'azquez-Semadeni, \& G\'omez 2012, and Fig.~1 of Zamora-Avil\'es \& V\'azquez-Semadeni 2014).

Finally, while it is possible that, in order to form the most massive fragments, additional compression mechanisms besides gravity may be necessary, the bulk of the fragmentation process in cores actively forming stars seems to be dominated by gravity rather than by turbulence. This is consistent with recent claims that the bulk of the non-thermal motions in clouds and cores may be dominated by infall rather than by random turbulence (\eg\ V\'azquez-Semadeni et al. 2008, Schneider et al. 2010, Ballesteros-Paredes et al. 2011a, Peretto et al. 2013, Gonz\'alez-Samaniego et al. 2014; see also the review by V\'azquez-Semadeni 2015).


\subsection{Physical implications}  
  
Our results suggest that \Nmm\ does not seem to depend significantly on the internal supersonic motions of the core, and are thus contrary to the widespread notion that support against gravity is necessary and that turbulence is the main physical process providing it and causing the fragmentation of molecular clouds. 
Since non-thermal supersonic motions are indeed observed in massive dense cores, but 
they do not seem to be random enough to act as a pressure against gravity, we propose that the observed `turbulence' cannot be used to define a `turbulent-Jeans' mass. 

Although turbulence may very well play a crucial role in the formation of the seeds of what eventually will grow as cores, as
demonstrated by the early evolution of molecular clouds in numerical simulations (e.g., Clark \& Bonnell 2005; V\'azquez-Semadeni et al. 2007, Heitsch \& Hartmann 2008), one possible interpretation of our results is that the fragmentation process in star-forming regions is controlled mainly by gravitational contraction and the ensuing reduction in the thermal Jeans mass as the density increases during the collapse. 
Thus, a possibility is that the non-thermal motions are dominated by infall, produced by the gravitational contraction (\eg\ Ballesteros-Paredes et al. 2011a). 
Indeed, analysis of the dense regions in simulations of driven, isothermal turbulence, indicate that the overdensities tend to have velocity fields with a net negative divergence (i.e. a convergence), rather than being completely random with zero or positive net divergence, as would be necessary for the bulk motions to exert a `turbulent pressure' capable of opposing the self-gravity of the overdensities (V{\'a}zquez-Semadeni, et al. 2008, Gonz{\'a}lez-Samaniego et al. 2014). Another possibility is that the non-thermal motions are strongly affected by stellar feedback, but in this case the effect on the clouds may be disruptive rather than supportive (Col\'in et al. 2013).

Our work supports the notion that non-thermal motions cannot be treated as capable of exerting a net turbulent
pressure that can provide support against gravity and stabilize the cores, since we have found no evidence that the `turbulent Jeans mass' plays any significant role in the fragmentation of the cores. If this view is correct, then theoretical models based on the hypothesis of turbulent support (\eg\ McKee \& Tan 2003; Krumholz \& McKee 2005) should be revised, as well as observational works that oversimplify the role of turbulence and estimate the Jeans mass by using an `effective sound speed' corresponding to the non-thermal velocity dispersion.  Clearly, a detailed comparison with simulations is needed to understand the origin of the non-thermal motions in massive dense cores and their role in the fragmentation process of molecular clouds.
These simulations would help establish more clearly that gravity is indeed controlling fragmentation in massive dense cores at 0.1~pc scales, and even at scales of the entire molecular cloud once the clouds are well developed, as suggested by several authors (\eg\ Clark \& Bonnell 2005; V\'azquez-Semadeni et al. 2007).



\section*{Acknowledgements}

The authors are grateful to the anonymous referee for providing comments improving the clarity and quality of the paper.
We wish to acknowledge useful and enjoyable discussions with Qizhou Zhang. 
A.P.\ is grateful to Jennifer Wiseman for sharing the \nh\ data, to Asunci\'on Fuente, Tom\'as Alonso-Albi, Ke'nichi Tatematsu, and Sylvain Bontemps for sharing the \nth\ data, and to Tim Jenness for sharing the James Clerk Maxwell Telescope data of IRAS 22198+6336 (Jenness et al. 1995).
A.P. and L.A.Z. acknowledge financial support from UNAM-DGAPA-PAPIIT IA102815 grant, and CONACyT, M\'exico.
J.B.P. thanks financial support from UNAM-PAPIIT grant number IN103012.
\'A.S.-M. acknowledges support by the collaborative research project SFB 956, funded by the Deutsche Forschungsgemeinschaft (DFG).
R.E. is partially supported by the Spanish MICINN grant AYA2011-30228-C03. 
G.B. is supported by the Spanish MICINN grant AYA2011-30228-C03-01 (co-funded with FEDER funds).
M. F. acknowledges the hospitality of the Aspen Center for Physics, which is supported in part by the US National Science Foundation under grant PHY-1066293.
L. G. receives support from the Center of Excellence in Astrophysics and Associated Technologies (PFB-06), CONICYT (Chile) and CSIRO Astronomy and Space Science (Australia).

{}

\begin{appendix}

\section{Full names and coordinates of the sample}

In Table~A1 we provide the full names or other names, coordinates and distances for the sources in our sample.

\begin{table*}
\caption{Main properties of the sample of massive dense cores studied in this work}
\begin{center}
{\small
\begin{tabular}{llcc ccc}
\noalign{\smallskip}
\hline\noalign{\smallskip}
Short
&Full
&\multicolumn{2}{c}{Position\supa}
&Distance
&$L_\mathrm{bol}$\supb
&$M_\mathrm{obs}$\supc
\\
name
&name
&$\alpha (\rm J2000)$
&$\delta (\rm J2000)$
&(kpc)
&(\lo)
&(\mo)
\\
\noalign{\smallskip}
\hline\noalign{\smallskip}
1-IC1396N       		&IRAS\,21391+5802		&21:40:41.71   	&+58:16:12.8		&0.75	&290	&78\\
2-I22198		       	&IRAS\,22198+6336		&22:21:26.78   	&+63:51:37.6		&0.76	&340	&115\\
3-N2071-1		&NGC\,2071			&05:47:04.78   &+00:21:43.1	   	&0.42	&440	&80\\
4-N7129-2		&NGC\,7129-FIRS2		&21:43:01.68   	&+66:03:23.6	    	&1.25	&460	&81\\
5-CB3-mm		&CB3-mm				&00:28:42.70   	&+56:42:06.8		&2.50	&700	&169\\
6-I22172N      		&IRAS\,22172+	5549-N	&22:19:08.60   	&+56:05:02.0     	&2.40	&830	&119\\
7-OMC-1S      		&OMC-1S				&05:35:14.00   	&$-$05:24:00.0 	&0.45	&2000	&158\\
8-A5142         		&AFGL\,5142			&05:30:48.02   	&+33:47:54.5	  	&1.80	&2200	&356\\
9-I05358NE		&IRAS\,05358+3543-NE	&05:39:13.07   	&+35:45:50.5	     	&1.80	&3100	&1480\\
10-I20126			&IRAS\,20126+4104		&20:14:26.04   &+41:13:32.5	    	&1.64	&8900	&68\\
11-I22134      		&IRAS\,22134+5834		&22:15:09.23   &+58:49:08.9	       	&2.60	&11800	&222\\ 
12-HH80-81         	&IRAS\,18162$-$2048 	&18:19:12.10   	&$-$20:47:30.0	      	&1.70	&21900	&333\\ 	
13-W3IRS5          	&W3-IRS5			&02:25:40.77   	&+62:05:52.5		&1.95	&140000	&971\\
14-A2591			&AFGL\,2591			&20:29:24.90   	&+40:11:19.5		&3.00	&190000	&784\\
\hline	
15-Cyg-N53		&Cygnus\,X-N53		&20:39:03.10   	&+42:25:50.0		&1.40	&300	&675\\
16-Cyg-N12		&Cygnus\,X-N12		&20:36:57.40   &+42:11:27.5		&1.40	&320	&622\\
17-Cyg-N63		&Cygnus\,X-N63		&20:40:05.20   &+41:32:12.0		&1.40	&470	&160\\
18-Cyg-N48		&Cygnus\,X-N48		&20:39:01.50   	&+42:22:04.0		&1.40	&4400	&865\\
19-DR21-OH		&DR21-OH			&20:39:01.00   &+42:22:46.0		&1.40	&10000	&526\\
\hline
\end{tabular}
\begin{list}{}{}
\item[$^\mathrm{a}$] Approximate position of the center of the field of view (corresponding to a region of $\sim0.1$~pc of diameter) where fragmentation was assessed in Palau et al. (2013, 2014).
\item[$^\mathrm{b}$] Bolometric luminosity as given in Table 1 of Palau et al. (2014).
\item[$^\mathrm{c}$] $M_\mathrm{obs}$ is the mass computed analytically from the model of Palau et al. (2014), integrating until the radius where the density profile could be measured for each source. Note that for I22198 we present here an updated version of the model (see Appendix C).
\end{list}
}
\end{center}
\label{tregions}
\end{table*}

\section{\nh\ and \nth\ spectra}

\begin{figure*}
\begin{center}
\begin{tabular}[b]{c}
     \epsfig{file=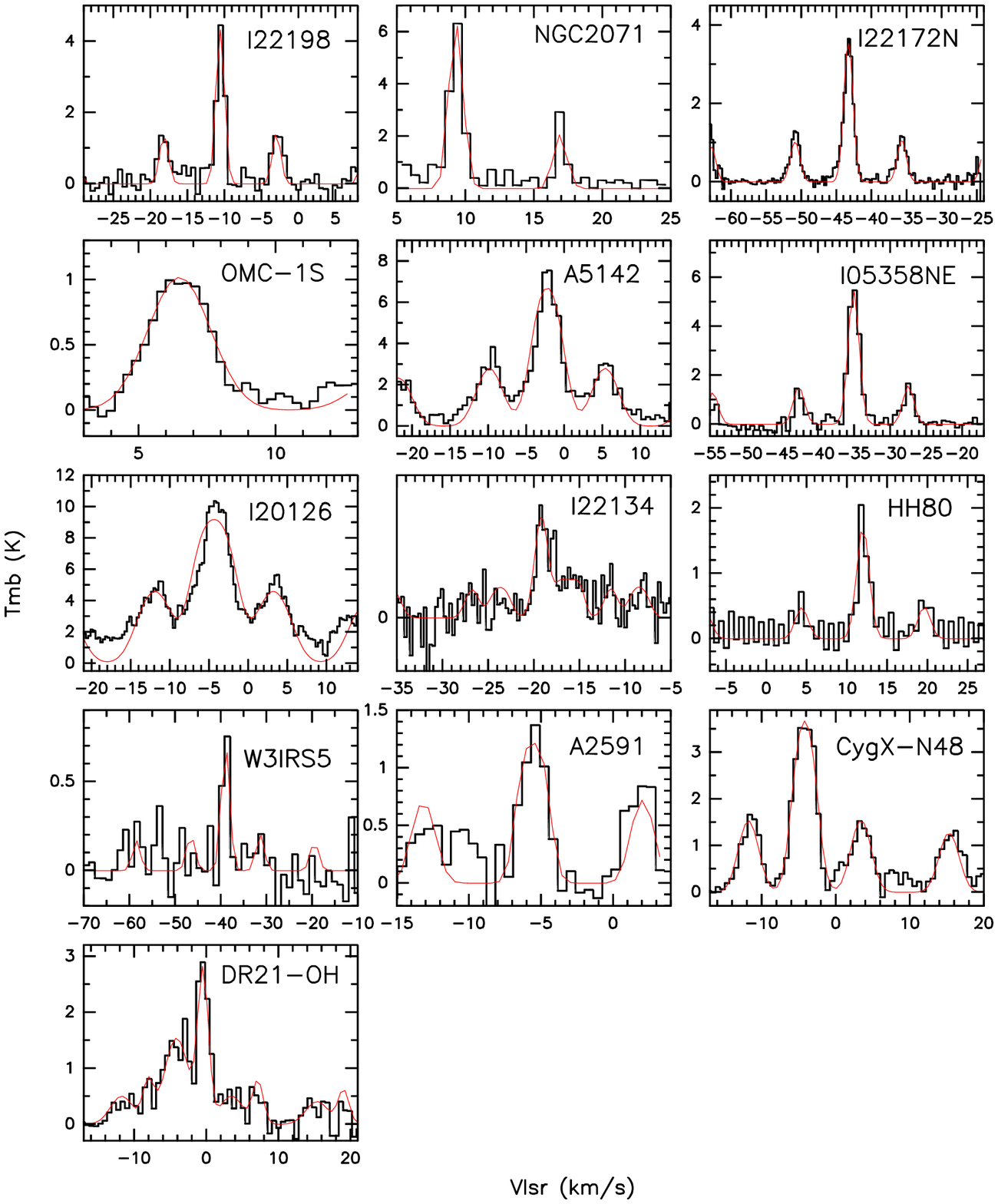, width=12.0cm,angle=0} \\
\end{tabular}
\caption{VLA \nh(1,1) spectra averaged over a region of $\sim0.1$~pc of diameter, where the fragmentation has been assessed. Red lines correspond to the CLASS fits to the hyperfine structure, from which the line width given in Table~1 has been inferred. For regions I22134 and DR21-OH, two velocity components have been used.}   
\label{fnh3spec}
\end{center}
\end{figure*}

\begin{figure*}
\begin{center}
\begin{tabular}[b]{c}
     \epsfig{file=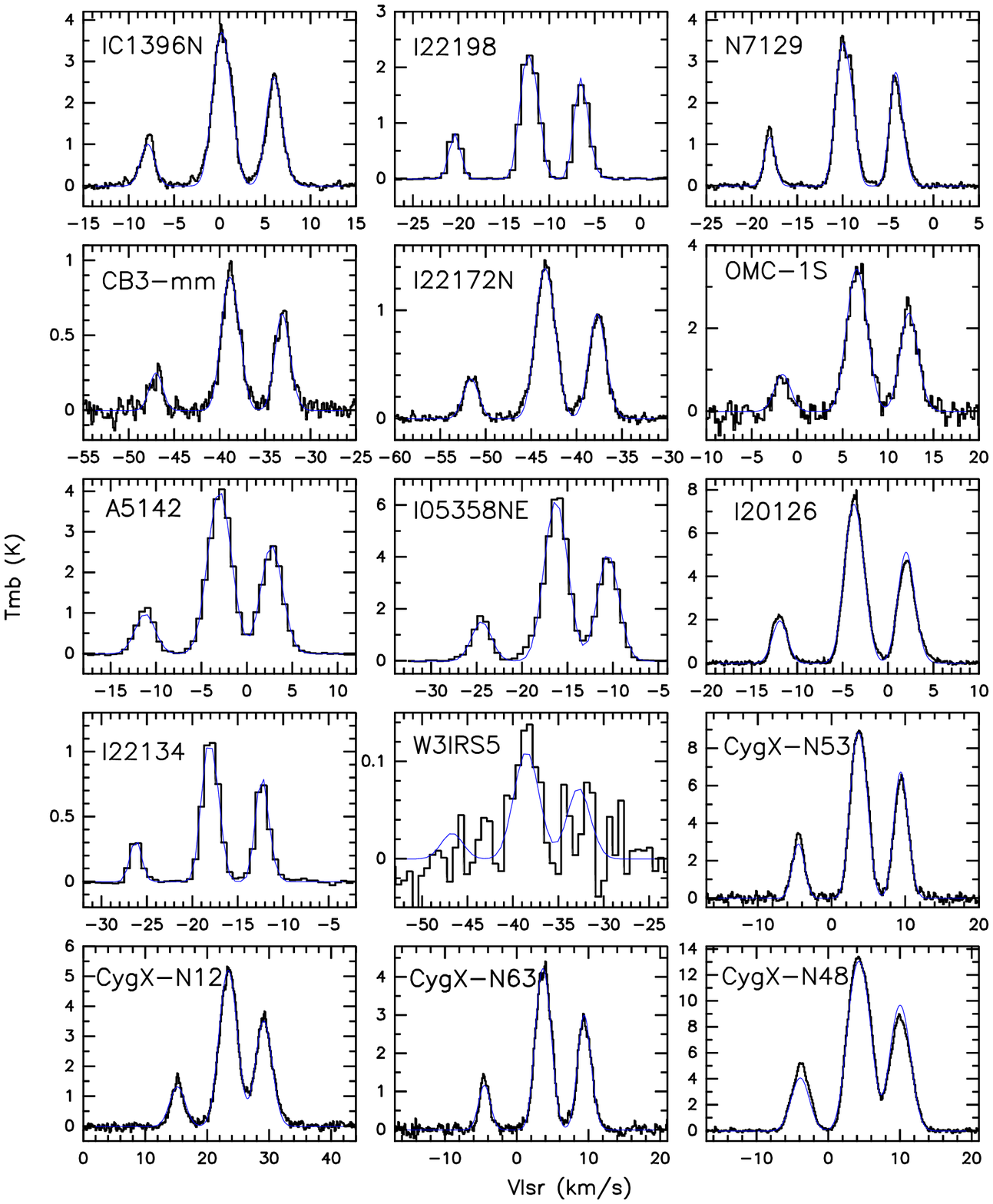, width=12.0cm,angle=0} \\
\end{tabular}
\caption{Single-dish \nth(1--0) spectra averaged over a region of $\sim0.1$~pc of diameter, where the fragmentation has been assessed. Blue lines correspond to the CLASS fits to the hyperfine structure, from which the line width given in Table~1 has been inferred.
}   
\label{fn2hspec}
\end{center}
\end{figure*}

In this appendix, we present the \nh(1,1) and \nth(1--0) spectra used in this work to estimate the non-thermal line widths for each massive dense core. The spectra, together with the hyperfine fits done using the CLASS program of GILDAS, are presented in Figs.~\ref{fnh3spec} and \ref{fn2hspec} and result from a compilation of data already published (see Section 2 for references) or reduced from the VLA archives.

\section{New density and temperature determination for IRAS\,22198+6336}

\begin{figure*}
\begin{center}
\begin{tabular}[b]{c}
     \epsfig{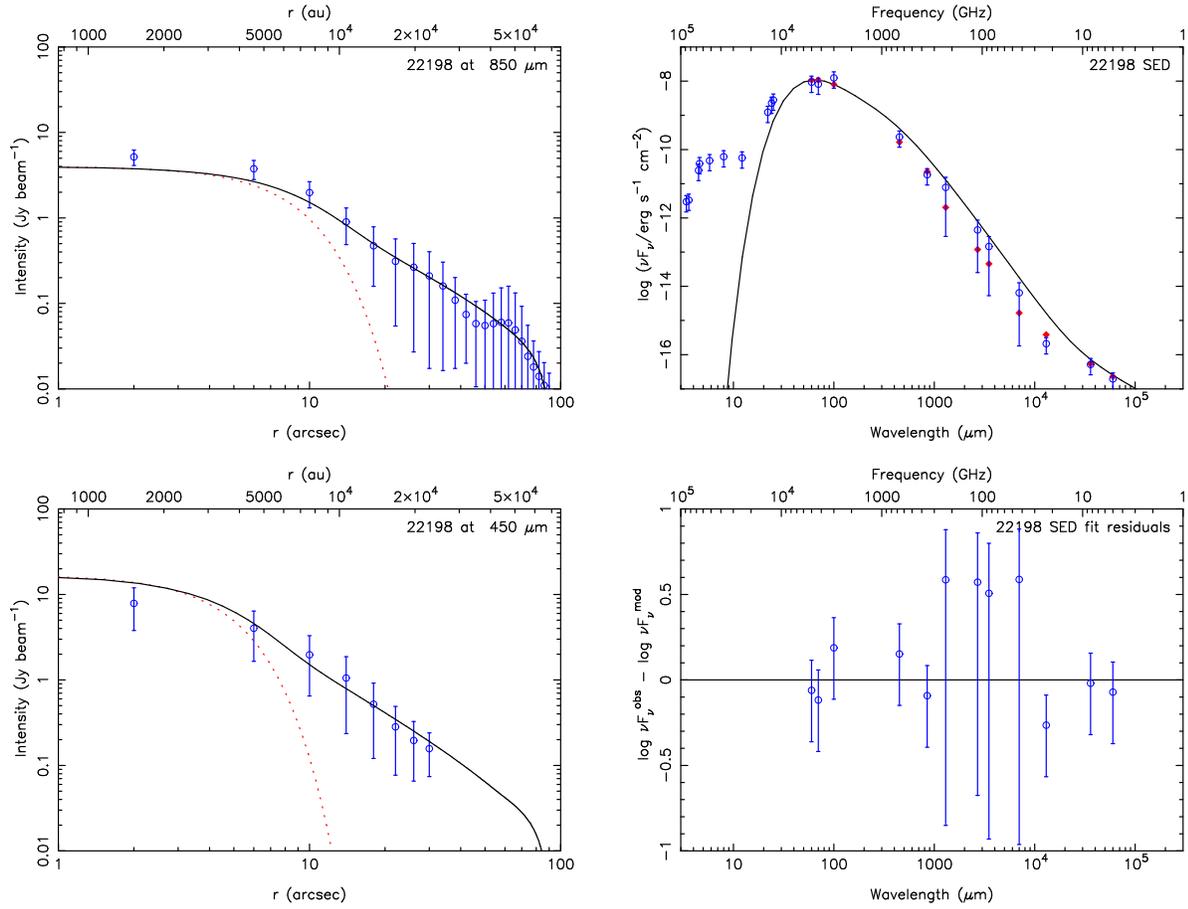} \\
\end{tabular}
\caption{New fit (after Palau et al. 2014) to the radial intensity profiles and Spectral Energy Distribution of IRAS\,22198+6336 (I22198) after using original JCMT data of Jenness et al. (1995) work.}   
\label{f22198}
\end{center}
\end{figure*}

In Palau et al. (2014) the original images of IRAS\,22198+6336 (I22198) published by Jenness et al. (1995) using the JCMT were not available, and the images were digitized. The images are now available and we have re-done the fit, with the additional difference (with respect to Palau et al. 2014) that the main beams assumed here are 7 and 14 arcsec at 450 and 850~$\mu$m, respectively, with no consideration of error beams. The results are presented in Fig.~\ref{f22198} and Table~\ref{t22198}.

\begin{table*}
\caption{Best-fit parameters to the radial intensity profiles and Spectral Energy Distribution of IRAS\,22198+6336 (I22198), and inferred properties (updated after Palau et al. 2014)}
\begin{center}
{\small
\begin{tabular}{lcr ccc ccc ccc}
\noalign{\smallskip}
\hline\noalign{\smallskip}
&
&$T_0$\supa
&$\rho_0$\supa
&
&
&
&$r_\mathrm{10\,K}$\supb
&$r_\mathrm{max}$\supb
&$M_\mathrm{obs}$\supb
&$\Sigma_\mathrm{0.1\,pc}$\supb
&$n_\mathrm{0.1\,pc}$\supb
\\
Source
&$\beta$\supa
&(K)
&($\times10^{-17}$ g cm$^{-3}$)
&$p$\supa
&$\chi_r$\supa
&$q$\supb
&(pc)
&(pc)
&(\mo)
&(g\,cm$^{-2}$)
&($\times 10^5$\,cm$^{-3}$)
\\
\noalign{\smallskip}
\hline\noalign{\smallskip}
I22198        &$1.16\pm0.22$	&$44\pm4$	&$3.4\pm0.4$	&$1.75\pm0.03$ &0.580 &0.39 	&0.22	&0.31	&115	&0.29	&3.6\\
\hline
\end{tabular}
\begin{list}{}{}
\item[$^\mathrm{a}$] Free parameter fitted by the model: $\beta$ is the dust emissivity index; $T_0$ and $\rho_0$ are the temperature and density at the reference radius, 1000 AU; $p$ is  the density power law index; $\chi_\mathrm{r}$ is the reduced $\chi$ as defined in equation (6) of Palau et al. (2014).
\item[$^\mathrm{b}$] Parameters inferred (not fitted) from the modeling. $q$ is the temperature power-law index, and $r_\mathrm{10\,K}$ is the radius of the core where the temperature has dropped down to $\sim10$~K; $r_\mathrm{max}$ is the radius at the assumed `ambient' density of 5000~\cmt; $M_\mathrm{obs}$ is the mass computed analytically from the model integrating until the radius where the density profile could be measured for each source. $\Sigma_\mathrm{0.1\,pc}$ and \ndiam\ are the surface density and density inside a region of 0.1~pc of diameter.
\end{list}
}
\end{center}
\label{t22198}
\end{table*}

\end{appendix}

\end{document}